\documentstyle[preprint,revtex]{aps} 
\begin{document}
\draft
\preprint{RU95--7--B}
\preprint{December 1995}
\begin{title}
Ground state and excitations of the extended Hubbard model
\end{title}
\author{D. F. Wang}
\begin{instit} 
Institut de Physique Th\'eorique\\
Ecole Polytechnique F\'ed\'erale de Lausanne\\
PHB-Ecublens, CH-1015 Lausanne, Switzerland
\end{instit}
\author{James T. Liu}
\begin{instit}
Department of Physics\\
The Rockefeller University\\
1230 York Avenue\\
New York, NY 10021--6399
\end{instit}
\begin{abstract}
We examine the ground state and excitations of the one dimensional extended 
Hubbard model with long range interaction.  The ground state wavefunction
and low lying excitations are given explicitly in the form of a Jastrow
product of two-body terms.  This result motivates an asymptotic Bethe ansatz
solution for the model.  We present evidence that this solution is in fact
exact and spans the complete spectrum of states.
\end{abstract}

\narrowtext

Low dimensional strongly correlated electron systems are of considerable
recent interest as possible models of high-$T_c$ superconductivity.  One
dimensional versions of such models have proven to be especially interesting
as they are often completely integrable and may be solved exactly.  A
particular case is the one dimensional Hubbard model, which was solved by
Lieb and Wu in 1967\cite{wu,yangnba}, and which exhibits a hidden $SO(4)$
symmetry \cite{yang1,yang2}. Recently, an extended Hubbard model was
introduced which is solvable by Bethe ansatz in one dimension
\cite{korepin2,exact}.  In this letter, we discuss the long range version
of the one dimensional extended Hubbard model on a uniform lattice.

There has been renewed interest in the integrable electronic models of 
Calogero-Sutherland type \cite{sutherland} of strong correlations since
the independent work of Haldane and Shastry \cite{haldane,shastry} on the
one dimensional spin chain of $1/r^2$ exchange interaction
\cite{kura,haldane2,shastry2,kawa,wang3,gebhard,fowler,poly,wang1,wang2}.
The system we consider here is the
Jastrow-wavefunction-integrable extended Hubbard model defined on a
one dimensional closed uniform chain of size $L$.  This model is a
generalization of the Haldane-Shastry spin chain and the long range
supersymmetric $t$-$J$ model and is similarly completely integrable.
In the following, we will investigate the ground state and excitations of 
the system in detail.
 
In the extended Hubbard model, each lattice site may be either unoccupied
(hole), occupied by a single electron (up-spin or down-spin) or doubly
occupied.  The latter case of localized electron pairs are denoted localons.
Thus there are four possible states at each site --- two bosonic and two
fermionic.  The Hamiltonian then takes the form
\begin{equation}
H=-{1\over 2} \sum_{1\le i\ne j\le L} J_{ij} \Pi_{ij}, 
\label{eq:hamil}
\end{equation}
where the permutation operator $\Pi_{ij}$  exchanges the states
between sites $i$ and $j$.  $\Pi_{ij}$ is given explicitly in terms of
electron operators by
\begin{eqnarray}
\Pi_{ij}=&&c_{j\uparrow}^\dagger c_{i\uparrow}^{\vphantom{\dagger}}
        (1-n_{i\downarrow}-n_{j\downarrow} )
+ c_{i\uparrow}^\dagger c_{j\uparrow}^{\vphantom{\dagger}} 
        (1-n_{i\downarrow}-n_{j\downarrow} )\nonumber\\
&&+ c_{j\downarrow}^\dagger c_{i\downarrow}^{\vphantom{\dagger}}
        (1-n_{i\uparrow}-n_{j\uparrow})
+ c_{i\downarrow}^\dagger c_{j\downarrow}^{\vphantom{\dagger}} 
        (1-n_{i\uparrow}-n_{j\uparrow})\nonumber\\
&&+{1\over 2} (n_i-1)(n_j-1)
+ c_{i\uparrow}^\dagger c_{i\downarrow}^\dagger
        c_{j\downarrow}^{\vphantom{\dagger}} c_{j\uparrow}^{\vphantom{\dagger}}
+c_{i\downarrow}^{\vphantom{\dagger}} c_{i\uparrow}^{\vphantom{\dagger}} 
        c_{j\uparrow}^\dagger c_{j\downarrow}^\dagger\nonumber\\
&&-{1\over 2} (n_{i\uparrow} - n_{i\downarrow}) 
        (n_{j\uparrow} - n_{j\downarrow})\nonumber\\
&&-c_{i\downarrow}^\dagger c_{i\uparrow}^{\vphantom{\dagger}}
        c_{j\uparrow}^\dagger c_{j\downarrow}^{\vphantom{\dagger}}
-c_{i\uparrow}^\dagger c_{i\downarrow}^{\vphantom{\dagger}}
        c_{j\downarrow}^\dagger c_{j\uparrow}^{\vphantom{\dagger}}\nonumber\\
&&+ (n_{i\uparrow} -1/2) (n_{i\downarrow} - 1/2) 
+ (n_{j\uparrow} -1/2) (n_{j\downarrow} -1/2).
\label{eq:permu}
\end{eqnarray}
The operators $c_{i\sigma}^{\vphantom{\dagger}}$ and $c_{i\sigma}^\dagger$
are electron annihilation and creation operators at site $i$ with spin
$\sigma=\uparrow, \downarrow$. The electron number operators are given by 
$n_{i\uparrow} = c_{i\uparrow}^\dagger c_{i\uparrow}^{\vphantom{\dagger}}$, 
$n_{i\downarrow}=c_{i\downarrow}^\dagger c_{i\downarrow}^{\vphantom{\dagger}}$
and $n_i=n_{i\uparrow} +n_{i\downarrow}$.

Because of the fermionic nature of the electrons, $\Pi_{ij}$ picks up an
appropriate sign when exchanging fermions, and hence acts as a graded
permutation operator.  This leads to an $SU(2|2)$ supersymmetry of the
system \cite{korepin2}.  In fact there is a natural generalization of this
model to a chain where there are $m$ bosonic and $n$ fermionic species at
each site, leading to an $SU(m|n)$ supersymmetry.  In this language, the
Haldane-Shastry spin chain is of type $SU(0|2)$ and the $t$-$J$ model is
of type $SU(1|2)$.  While the following results are easily generalized to
the $SU(m|n)$ case, we are solely concerned with the $SU(2|2)$ extended
Hubbard model in this letter.

In addition to the grading of the permutation operator, the coupling
parameter $J_{ij}$ determines the nature of this model.  The extended
Hubbard model studied by E{\ss}ler, Korepin and Schoutens
\cite{korepin2} corresponds to nearest neighbor interactions,
$J_{ij} =\delta_{1,|i-j|}$.  We are interested instead in the long range
version of this model where $J_{ij}=1/d^2(i-j)$, and
$d(n)={L\over \pi} \sin(\pi n/L)$.

Since the Hamiltonian given by Eq.~(\ref{eq:hamil}) acts only as
permutations, the total numbers of each species are all conserved
quantities.  We denote the number of localons, unpaired up-spin electrons,
unpaired down-spin electrons and holes by
\begin{eqnarray}
&&\bar A=\sum_{i=1}^L n_{i\downarrow} n_{i\uparrow}\nonumber\\
&&\bar M_\uparrow=\sum_{i=1}^L n_{i\uparrow}-\bar A\nonumber\\ 
&&\bar M_\downarrow=\sum_{i=1}^L n_{i\downarrow}-\bar A\nonumber\\
&&\bar Q=L-\bar M_\uparrow - \bar M_\downarrow - \bar A.
\end{eqnarray}
The total number of unpaired electrons may then be written as
$\bar M=\bar M_\uparrow+ \bar M_\downarrow$.
In the following, we work in the Hilbert space of fixed 
$\bar A, \bar M_\uparrow, \bar M_\downarrow$, and $\bar Q$. 
The system reduces to the $SU(1|2)$ long range $t$-$J$ model when $\bar A=0$
and the $SU(0|2)$ Haldane-Shastry spin chain in the limiting case when
$\bar Q=0$ in addition to $\bar A=0$.

In order to construct the ground state and excitations, we write the
states of the system in terms of excitations from the fully
polarized up-spin state, $|F\rangle=\prod_{i=1}^L c_{i\uparrow}^\dagger
|0\rangle$.  Thus a general state vector is given by
\begin{equation}
|\phi\rangle =\sum_{x, y, z} \phi(x,y,z) \prod_{i=1}^{\bar A} 
c_{x_i\downarrow}^\dagger \prod_{\alpha=1}^{\bar M_\downarrow}
c_{y_\alpha\downarrow}^\dagger c_{y_\alpha\uparrow}^{\vphantom{\dagger}}
\prod_{l=1}^{\bar Q} c_{z_l\uparrow}^{\vphantom{\dagger}} |F\rangle.
\end{equation}
The amplitude $\phi$ is antisymmetric in the positions of the localons,
$x=(x_1, x_2, \cdots, x_{\bar A})$, symmetric in the positions of the
unpaired down-spins, $y=(y_1, y_2, \cdots, y_{\bar M_\downarrow})$, 
and antisymmetric in the positions of the the holes,
$z=(z_1, z_2, \cdots, z_{\bar Q})$.  When the state vector is written
in the above manner, the singly-occupied up-spin states correspond to the
sites unoccupied by the localons, holes, and singly-occupied down-spin
electrons.  The fact that all four species completely fill the lattice
has proven useful in allowing alternate equivalent representations in terms
of excitations from uniform backgrounds of any of the three other species
\cite{wang3}.

Assuming that $S_z=(\bar M_\uparrow- \bar M_\downarrow)/2\ge 0$, 
we propose the following Jastrow wavefunction
\begin{eqnarray}
\phi=&&e^{{2i\pi \over L}
[J_h(\sum_i x_i+\sum_l z_l) + J_s \sum_\alpha y_\alpha]}
\times \prod_{1\le \alpha< \beta \le \bar M_\downarrow} d^2(y_\alpha-y_\beta)
\times\nonumber\\ 
&&\times \prod_{\alpha=1}^{\bar M_\downarrow} (\prod_{i=1}^{\bar A}
d(y_\alpha-x_i)
\cdot \prod_{l=1}^{\bar Q} d(y_\alpha-z_l)) \times
\prod_{1\le i< j \le \bar A} d(x_i-x_j) \times\nonumber\\
&&\times \prod_{1\le l<n \le \bar Q}
d(z_l-z_n) \times \prod_{l,i} d(x_i-z_l),
\label{eq:wavefunction} 
\end{eqnarray} 
where the quantum numbers $J_h$ and $J_s$ are either integers or
half-integers as appropriate so that the wavefunction is single-valued
under any of the individual translation operations $x_i\rightarrow x_i+L$,
$y_\alpha\rightarrow y_\alpha+L$ or $z_l\rightarrow z_l+L$.  The Jastrow
wavefunction is constructed to satisfy the appropriate symmetries upon
interchange of any pair of sites.

Inspection of the wavefunction indicates that the localons and the holes
play the same role.  This may be understood from the $SU(2)$ symmetry relating
the two bosonic states.  Furthermore, Sutherland has shown that the ground
state energy for a general $SU(m|n)$ permutation model coincides with that of
the equivalent $SU(1|n)$ model \cite{sutherland2}.  We therefore can
take identical localon and hole momenta, $J_h$, and can then compute the
effects of the Hamiltonian acting on the wavefunction as if one were dealing
with the $SU(1|2)$ supersymmetric $t$-$J$ model of $1/r^2$ hopping and
exchange.  The number of ``holes'' $Q$ in the effective $t$-$J$ model is given
by the total number of bosons, $Q=\bar Q+\bar A$.

Using this analogy to the $t$-$J$ model, we have computed the action of the
Hamiltonian $H$ on the Jastrow wavefunction defined by
Eq.~(\ref{eq:wavefunction}).  We find that the wavefunction is in fact an
exact eigenstate of the system provided the quantum numbers $J_s$ and $J_h$
satisfy the following constraints:
\begin{eqnarray}
|J_h-L/2|&\le&(\bar M_\uparrow+1)/2 \nonumber\\ 
|J_s-L/2|&\le&1+S_z/2\nonumber\\                
|J_s-J_h|\,\,&\le&(\bar M_\downarrow+1)/2.
\end{eqnarray}
Under these conditions, the energy of this state, as a function of the
quantum numbers $J_h$ and $J_s$, is given by
\begin{eqnarray}
E/(\pi^2/L^2)=&&{1\over6}L(L^2-1)+{1\over2}Q(\bar M_\downarrow+Q)
                                  (2\bar M_\downarrow-Q)\nonumber\\
              &&+\bar M_\downarrow[{2\over3}(\bar M_\downarrow^2-1)
                                    -2 J_s(L-J_s)]\nonumber\\
              &&+\,Q\,\,[{2\over3}(Q^2\,-1)-2J_h(L-J_h)]\nonumber\\
              &&+\,2Q(J_s-J_h)^2. 
\end{eqnarray}

We recall that the above eigenenergies have been calculated in a sector
of fixed $\bar M_\uparrow$, $\bar M_\downarrow$, $\bar Q$, and $\bar A$.
The ground state of the Hamiltonian in the Hilbert space of fixed $\bar Q$
and $\bar A$ is then given by the Jastrow wavefunction in the $S_z=0$ or
$1/2$ sector with $J_s$ and $J_h$ both as close to $L/2$ as possible.
Since the ground state energy is identical among all the $\bar Q+\bar A+1$
states with total boson number $Q$, the ground state falls in either the
$({\bf 1},\bar Q+\bar A+1)$ or the $({\bf 2},\bar Q+\bar A+1)$ representation
of $SU(2)\times SU(2)\subset SU(2|2)$ for $S_z=0$ or $S_z=1/2$ respectively.
As a consequence, the lowest energy state in the complete Hilbert space is
given by $Q=L$ and is $L+1$ fold degenerate. The corresponding state vectors 
are given by $|G\rangle =\sum_{\{x\}} \prod_{i=1}^{\bar A}
c_{x_i\uparrow}^\dagger c_{x_i\downarrow}^\dagger |0\rangle$, 
where $\bar A=0, 1, 2, \ldots, L$.  Any of these ground states with
$\bar A> 0$ is superconducting and exhibits off-diagonal long range order.
The state vectors are very similar to the $\eta$-pairing eigenstates for
the conventional Hubbard model constructed by Yang\cite{yang2}, except
that the phase of each localon is 1.   


The Jastrow wavefunctions corresponding to other
values of $J_s$ and $J_h$ will be excited states of the system.
One may compute the corresponding spin and charge excitation energies
relative to the ground state.  In particular, for the Jastrow wavefunctions
written in this way, various correlation functions, such as the correlation
functions between the localons, the holes, and the singly-occupied
down spins, can be computed in a compact form following the recent work of
Forrester \cite{forrester}.

For the extended Hubbard model, the above wavefunctions do not span the
complete energy spectrum of the system.  This comes as no surprise since
the Jastrow product only describes uniform excitations of the various
particles.  Nevertheless, since we have shown the ground state of this model
to be a product of two-body functions, we are led to a solution in the form
of an asymptotic Bethe ansatz (ABA) \cite{sutherland,kawa}.  With two
bosonic and two fermionic species to consider, the nesting of the Bethe ansatz
may be performed in various manners \cite{yangnba,suthnba,sutherland2,exact}.
We choose the BBFF grading and start from the fully polarized background
$|F\rangle$.  Introducing three sets of pseudomomenta,
$\{p_\alpha\}$, $\{q_l^{(1)}\}$ and $\{q_i^{(2)}\}$, the ABA equations are
of the form
\begin{eqnarray}
p_\alpha L&=&2\pi J_\alpha+\pi\!\!\!\!\!\!
\sum_{\beta=1}^{\bar M_\downarrow+\bar Q+\bar A}\!\!\!\!\!
{\rm sgn}(p_\alpha-p_\beta)
-\pi\!\sum_{l=1}^{\bar Q+\bar A}{\rm sgn}(p_\alpha-q_l^{(1)}) \nonumber\\
2\pi I_l^{(1)}&=&\pi\!\!\!\!\!\!
\sum_{\alpha=1}^{\bar M_\downarrow+\bar Q+\bar A}
\!\!\!\!\!{\rm sgn}(q_l^{(1)}-p_\alpha)
-\pi\sum_{i=1}^{\bar A}{\rm sgn}(q_l^{(1)}-q_i^{(2)}) \nonumber\\
2\pi I_i^{(2)}&=&\pi\!\sum_{l=1}^{\bar Q+\bar A}{\rm sgn}(q_i^{(2)}-q_l^{(1)})
-\pi\sum_{j=1}^{\bar A}{\rm sgn}(q_i^{(2)}-q_j^{(2)}).
\label{eq:aba}
\end{eqnarray}
The resulting energy and momentum depend only on the $\{p_\alpha\}$ and are
given by
\begin{eqnarray}
E&=&{1\over2}\sum_{\alpha=1}^{\bar M_\downarrow+\bar Q+\bar A}
(p_\alpha^2-\pi^2),\nonumber\\
P&=&\sum_{\alpha=1}^{\bar M_\downarrow+\bar Q+\bar A}(p_\alpha-\pi), 
\end{eqnarray}
where the reference energy of the system is ${\pi\over6}L(1-{1\over L^2})$,
and the reference momentum of the system is $(L-1)\pi$.  

The (individually non-overlapping) quantum numbers $J_\alpha$, $I_l^{(1)}$
and $I_i^{(2)}$ are either integral or half-integral and characterize the
ABA state.  They are restricted to lie in the range
\begin{eqnarray}
|J_\alpha|&\le&(L-\bar M_\downarrow-1)/2\nonumber\\
|I_l^{(1)}|&\le&(\bar M_\downarrow+\bar Q-2)/2\nonumber\\
|I_i^{(2)}|&\le&(\bar Q-1)/2.
\end{eqnarray}
As a result, in this BBFF picture, the ABA equations only hold in the
sectors of the Hilbert space satisfying
$\bar M_\uparrow\ge \bar M_\downarrow\ge \bar A+1$ and
$\bar Q\ge \bar A$.  Empirical observations on small systems ($L\le10$)
indicate that the ABA generates highest-weight states of the $SU(2|2)$
superalgebra.  This fact may be used to construct entire states of
the extended Hubbard model.  However, simple counting rules based on
spinon degrees of freedom (zeros in $J_\alpha$) \cite{haldane2,wang3}
no longer generate the entire states when additional species are present.
In order to account for the complete states, it is important to modify
the ABA equations by taking the ``squeezed''-string picture into account
\cite{hahaldane}.  Since the resulting Ha-Haldane equations are similar to
the {\it exact} Bethe ansatz solution of the nearest neighbor model
\cite{exact}, the counting of states proceeds in a similar manner
\cite{korepin1,schoutens}.

In both the Haldane-Shastry model and the supersymmetric $t$-$J$
model, the ABA is known to give exact results, even in the non-asymptotic
limit \cite{sutherland,haldane2,wang3}.  While it would be a more
complicated procedure to prove the same for Eqs.~(\ref{eq:aba}), the
integrability of this model as well as its similarities to the $t$-$J$ model
should enable such a proof.

In summary, we have provided the ground state and uniform excitations of the
long range extended Hubbard model.  These wavefunctions are in the form of a
Jastrow product, for which various correlators are already available.
This system of strongly correlated electrons is completely integrable
with an infinite set of mutually commuting constants of motion.  This fact
enables us to determine the full energy spectrum in terms of an asymptotic
Bethe ansatz and its Ha-Haldane modification.  As a result, it should be
possible to analytically study the thermodynamics and elementary excitations
of this model.  Since we only needed to know the general properties of the
graded permutation operator $\Pi_{ij}$, our results are easily generalized
to the arbitrary long range $SU(m|n)$ model.

This work was supported in part by the Swiss National Science Foundation
and by the U.~S.~Department of Energy under grant no.~DOE-91ER40651-TASKB.

\end{document}